\documentclass[12pt]{article}

\begin{document}
\newcommand{\be}{\begin{equation}}
\newcommand{\ee}{\end{equation}}
\newcommand{\lll}{\lambda}
\def\Journal#1#2#3#4{{#1} {\bf #2}, #3 (#4)}

\newcommand{\A}{\alpha}
\newcommand{\B}{\beta}
\newcommand{\T}{\theta}
\newcommand{\Ep}{\epsilon}
\newcommand{\beq}{\begin{equation}}
\newcommand{\eeq}{\end{equation}}
\newcommand{\fr}{\frac}
\newcommand{\beqn}{\begin{eqnarray}}
\newcommand{\eeqn}{\end{eqnarray}}
\newcommand{\G}{\gamma}
\newcommand{\D}{\delta}
\renewcommand{\P}{\phi}
\newcommand{\intl}{\int\limits_{1}^{\infty}}
\renewcommand {\L}{\lambda}
\newcommand{\pt}{\partial}

\newcommand{\bq}{\bar q_A}
\newcommand{\tq}{\tilde q_A}
\newcommand{\btq}{\bar{\tilde q}^A}
\newcommand{\fa}{\varphi^A}
\newcommand{\bfa}{\bar \varphi_A}

\newcommand{\none}{{\cal N}=1}
\newcommand{\ntwo}{{\cal N}=2}
\newcommand{\nzer}{{\cal N}=0}

\def\st{\scriptstyle}
\def\sst{\scriptscriptstyle}
\def\mco{\multicolumn}
\def\epp{\epsilon^{\prime}}
\def\vep{\varepsilon}
\def\ra{\rightarrow}
\def\al{\alpha}
\def\ab{\bar{\alpha}}
\def\be{\begin{equation}}
\def\ee{\end{equation}}
\def\bea{\begin{eqnarray}}
\def\eea{\end{eqnarray}}
\def\CPbar{\hbox{{\rm CP}\hskip-1.80em{/}}}

\begin{flushright}
\begin{tabular}{l}
ITEP-TH-25/07\\
\end{tabular}
\end{flushright}

\vskip2.5cm

\centerline{\large
\bf Nonabelian strings  in a dense  matter }

\vspace{1.5cm}

\centerline{\sc  A. Gorsky$^a$ and V. Mikhailov$^{a,b}$}

\vspace{2.5mm}

\centerline{a) \it Institute of Theoretical and Experimental Physics }
\centerline{\it  B. Cheremushkinskaya ul. 25, 117259 Moscow, Russia}
\vspace{0.5 cm}
\centerline{b) \it Moscow Institute of Physics and Technology}
\centerline{\it Dolgoprudnyi, Moscow region, Russia}

\def\thefootnote{\fnsymbol{footnote}}%
\vspace{4.5cm}

\centerline{\large \bf Abstract}
\vspace{1.5cm}
We consider gauge theories  with scalar
matter with and without supersymmetry at nonzero chemical
potential. It is argued that a chemical potential plays a role
similar to the FI term. We analyze theory at weak coupling regime at large chemical potential and argue that it supports nonabelian
non-BPS strings.
Worldsheet theory on the nonabelian string in a dense
matter is briefly discussed.
\vspace{5mm}
\newpage
\section{Introduction}

Recent studies in QCD provide a rich phase
diagram involving color superconductivity \cite{arw} and
color-flavor locking phases at high density (see, for review
\cite{review}). High
density region due to the asymptotic freedom is treatable perturbatively and
it is natural to use this lucky possibility to the full extent.
In particular it is interesting to investigate
if the extended objects like strings
and domain walls  exist in the theory with
chemical potential. The example of such object whose very
existence is ultimately related to the presence of a
non-vanishing density has been found . Indeed the tension
of the domain wall discussed in \cite{son}(see also \cite{fz,kaplan}) is proportional
to the chemical potential hence it admits a quasiclassical description at high density.

On the other hand the gauge/string duality provides the
description of the SUSY gauge theory
at strong coupling regime. To discuss
the dense matter along this way it
is necessary to determine the background for the
closed string which would encode a non-vanishing chemical
potential. This background has been found in
\cite{ads} and it was demonstrated \cite{kim,ht}
that the gravity description clearly indicates
the deconfinement phase transition at large
enough chemical potential. Similar studies have
been done  within the holographic description
of  QCD-like theory in Sakai-Sugimoto model \cite{ss}.
Some other aspects of the
dense SUSY gauge theories  has been discussed in $\none$ \cite{n=1}
and $\ntwo$ \cite{n=2} cases.

In this letter we shall search for the extended
objects in a dense  matter with and without SUSY.
Our main goal shall be investigation of the
nonabelian strings found recently in softly broken
$\ntwo$ SYM theories
with fundamental matter \cite{auzzi,tong1}. They
exist at weak coupling regime
and therefore are under  theoretical control. The nonabelian
strings exist in  $\nzer$ \cite{gsy}
gauge theories with the scalar matter as well (see \cite{rev2} for the reviews).
Moreover there are strong indications that they provide the proper pattern for the explanation
of the nonperturbative  lattice QCD data
\cite{gz}.
In $\ntwo$ case nonabelian
strings are BPS objects keeping some amount of SUSY
which protects their stability. In theories with
less amount of SUSY this argument is lost nevertheless one
can discuss the nonabelian strings at weak coupling
regime on the firm ground. The theory on the string
worldsheet can be analyzed explicitly, in particular the
spectrum of the worldsheet theory involves kinks
or kink-antikink pairs which can be identified with monopoles
or monopole-antimonopole pairs from the 4d bulk viewpoint
\cite{tonmon,sy2004}.

The question under investigation in this letter concerns the existence
of nonabelian strings if the chemical potential is switched on.
It turns out that the classical nonabelian string solution does exist in a weakly
coupled dense  matter in the theory with
large Fayet-Iliopoulos (FI) term $\xi$  and/or chemical
potential $\mu$. Moreover strings still exist if we
switch off FI term at all. We shall consider strings
both in non-SUSY and SUSY theories and
argue that the worldsheet theory
on the nonabelian string is non-supersymmetric.
Similar to the previous analysis \cite{tonmon,sy2004}
it can be shown that at least at large $\xi$ and small
$\mu$ there are kinks
in the bosonic action on the nonabelian
string which can be identified with the monopoles from
the four-dimensional perspective. Since worldvolume
theory is non-supersymmetric kinks are in the
confining phase.

The paper is organized as follows. In Section 2 we discuss
the non-SUSY gauge model with the scalar matter. It is shown
that chemical potential $\mu$ for U(1) charge plays the role similar to the
FI term $\xi$ that is large $\mu$ limit supports the semiclassical
nonabelian strings. In Section 3 we consider the softly broken
SUSY models with the chemical potential. Section 4 concerns the comments on the worldsheet
theory and the last Section is devoted to the conclusion.

\section{Non-supersymmetric model}
Here we discuss the simplest model which can be used
to analyze nonabelian strings.   The gauge group of the model
is SU($N)\times$U(1). The model contains SU($N$) and U(1)
gauge bosons
and $N$ scalar fields charged with respect to
$U(1)$ which form $N$ fundamental representations of SU($N$).
It is convenient to write these fields in the form of
$N\times N$ matrix $\Phi =\{\varphi^{kA}\}$
where $k$ is the SU($N$) gauge index while $A$ is the flavor
index,
\beq
\Phi =\left(
\begin{array}{cccc}
\varphi^{11} & \varphi^{12}& ... & \varphi^{1N}\\[2mm]
\varphi^{21} & \varphi^{22}& ... & \varphi^{2N}\\[2mm]
.....&...&...&...\\[2mm]
\varphi^{N1} & \varphi^{N2}& ... & \varphi^{NN}
\end{array}
\right)\,.
\label{phima}
\eeq
The action of the model reads as\,
\beqn
S &=& \int {\rm d}^4x\left\{-\frac1{4g_2^2}
\left(F^{a}_{\mu\nu}\right)^{2}
- \frac1{4g_1^2}\left(F_{\mu\nu}\right)^{2}
 \right.
 \nonumber\\[3mm]
&+&
 {\rm Tr}\, (\nabla_\mu \Phi)^\dagger \,(\nabla^\mu \Phi )
-\frac{g^2_2}{2}\left[{\rm Tr}\,
\left(\Phi^\dagger T^a \Phi\right)\right]^2
 -
 \frac{g^2_1}{8}\left[ {\rm Tr}\,
\left( \Phi^\dagger \Phi \right)- N\xi \right]^2
 \nonumber\\[3mm]
 &+&\left.
 \frac{i\,\theta}{32\,\pi^2} \, F_{\mu\nu}^a \tilde{F}^{a\,\mu\nu}
 \right\}\,,
\label{redqed}
\eeqn
where $T^a$ stands for the generator of the gauge SU($N$),
\beq
\nabla_\mu \, \Phi \equiv  \left( \partial_\mu -\frac{i}{\sqrt{ 2N}}\; A_{\mu}
-i A^{a}_{\mu}\, T^a\right)\Phi\, ,
\label{dcde}
\eeq
and $\theta$ is the vacuum angle. The last
term in the second line
forces $\Phi$ to develop a vacuum expectation value (VEV) while the
last but one term
forces the VEV to be diagonal,
\beq
\Phi_{\rm vac} = \sqrt\xi\,{\rm diag}\, \{1,1,...,1\}\,.
\label{diagphi}
\eeq

We assume the FI parameter $\xi$ to be large,
\beq
\sqrt{\xi}\gg \Lambda_4,
\label{weakcoupling}
\eeq
where $\Lambda_4$ is the scale of the four-dimensional theory (\ref{redqed}).
This ensures the weak coupling regime as both couplings $g^2_1$ and $g^2_2$
are frozen at a large scale.
The  vacuum field (\ref{diagphi}) results in  the spontaneous
breaking of both gauge and flavor SU($N$)'s.
A diagonal global SU($N$) survives
\beq
{\rm U}(N)_{\rm gauge}\times {\rm SU}(N)_{\rm flavor}
\to {\rm SU}(N)_{\rm diag}\,.
\eeq
yielding color-flavor locking phase in the vacuum.

The inclusion of the chemical potential into the relativistic invariant
theory goes as follows \cite{kapusta}. The covariant derivative for scalars
gets
shifted by the term $i\mu\delta_{\nu 0}$ which yields the total potential in
the theory
\beq
V= \frac{g^2_2}{2}\left[{\rm Tr}\,
\left(\Phi^\dagger T^a \Phi\right)\right]^2
+\frac{g^2_1}{8}\left[ {\rm Tr}\,
\left( \Phi^\dagger \Phi \right)- N\xi \right]^2  - \mu^2 {\rm Tr}\,
\left( \Phi^\dagger \Phi \right)
\eeq
It is clear that chemical potential provides the same symmetry breaking pattern
as
FI term and at large chemical potential theory is at weak coupling regime hence we get the effective FI term
\beq
\xi_{eff}=\xi + \fr{4\mu^{2}}{Ng_1^{2}}
\eeq

The topological argument providing the stability
of the string involves the
combination of the $Z_N$ center of SU($N$) with the elements $\exp (2\pi i
k/N)\in$U(1)
A topologically stable string solution
possesses both windings, in SU($N$) and U(1). In other words,
\beq
\pi_1 \left({\rm SU}(N)\times {\rm U}(1)/ Z_N
\right)\neq 0\,.
\eeq
and this nontrivial topology amounts to winding
of just one element of $\Phi_{\rm vac}$,  for instance,
\beq
\Phi_{\rm string} = \sqrt{\xi_{eff}}\,{\rm diag} ( 1,1, ... ,e^{i\alpha (x)
})\,,
\quad x\to\infty \,.
\label{ansa}
\eeq
Such strings can be called elementary, and the ANO string can be viewed as a
bound state of
$N$ elementary strings.

In the discussion of the vacuum structure in the presence of chemical potential
it was implicitly assumed that $A_\mu=0$ in the vacuum. However this
assumption
would violate the Gauss law unless we add a source term $J_\mu A_\mu$ with
constant
background charge density $J_\mu=J_0 \D_{\mu 0}$ \cite{kapusta}. The necessary
value of the background charge density $J_0$ is determined by requiring
$A_\mu=0$ in the vacuum.
In the other way, we can induce a chemical potential without changing the
covariant derivatives
by adding a term $J_\mu A_\mu$ to
the Lagrangian and considering the induced vev $A_0=\mu N$ as a chemical
potential.
The value of $\mu$-induced is then determined from the Gauss law,
\beq
\mu\left(\xi+\fr{4\mu^2}{Ng_1^2}\right)=-\sqrt\fr{N}{2}J_0\,.
\eeq
In the following we shall use this second way of adding the chemical potential.

The nonabelian string in fact is the twisted $Z_N$ string
hence let us first describe $Z_N$ solution which
can be written as follows \cite{auzzi}:
\beqn
\Phi &=&
\left(
\begin{array}{cccc}
\phi(r) & 0& ... & 0\\[2mm]
.....&...&...&...\\[2mm]
0& ... & \phi(r)&  0\\[2mm]
0 & 0& ... & e^{i\alpha}\phi_{N}(r)
\end{array}
\right) ,
\nonumber\\[5mm]
A^{{\rm SU}(N)}_i &=&
\frac1N\left(
\begin{array}{cccc}
1 & ... & 0 & 0\\[2mm]
.....&...&...&...\\[2mm]
0&  ... & 1 & 0\\[2mm]
0 & 0& ... & -(N-1)
\end{array}
\right)\, \left( \pt_i \alpha \right) \left[ -1+f_{NA}(r)\right] ,
\nonumber\\[5mm]
A^{{\rm SU}(N)}_0 &=&
-\frac1N g_{NA}\left(
\begin{array}{cccc}
1 & ... & 0 & 0\\[2mm]
.....&...&...&...\\[2mm]
0&  ... & 1 & 0\\[2mm]
0 & 0& ... & -(N-1)
\end{array}
\right)\, ,
\nonumber\\[5mm]
A^{{\rm U}(1)}_i &=& \sqrt{\frac{2}{N}}\,
\left( \pt_i \alpha \right)\left[1-f(r)\right] ,\nonumber\\
A^{{\rm U}(1)}_0 &=& \sqrt\frac{2}{N}g\,,
\label{znstr}
\eeqn
where $i=1,2$ labels coordinates in the plane orthogonal to the string
axis and $r$ and $\alpha$ are the polar coordinates in this plane. The profile
functions $\phi(r)$ and  $\phi_N(r)$ determine the profiles of the scalar
fields,
while $f_{NA}(r),\, g_{NA}(r)$ and $f(r),\, g(r)$ determine the SU($N$) and
U(1) fields of the
string solutions, respectively.

The equations for profile functions follow from the equations of motion,

\beqn
&&\fr{1}{g_1^2}\left(g''+\fr{g'}{r}\right)=\fr{1}{N}\left((N-1)(g-g_{NA})\P^2+(g+(N-1)g_{NA})\P_N^2\right)+\sqrt\fr{N}{2}J_0\nonumber\\
&&\fr{1}{g_2^2}\left(g_{NA}''+\fr{g_{NA}'}{r}\right)=\fr{1}{N}\left(-(g-g_{NA})\P^2+(g+(N-1)g_{NA})\P_N^2\right)\nonumber\\
&&\fr{r}{g_1^2}\left(\fr{f'}{r}\right)'=\fr{1}{N}\left((N-1)(f-f_{NA})\P^2+(f+(N-1)f_{NA})\P_N^2\right)\nonumber\\
&&\fr{r}{g_2^2}\left(\fr{f_{NA}'}{r}\right)'=\fr{1}{N}\left(-(f-f_{NA})\P^2+(f+(N-1)f_{NA})\P_N^2\right)\nonumber\\
&&\P''+\fr{1}{r}\P'=\left[\fr{1}{N^2}\left(\fr{1}{r^2}(f-f_{NA})^2-(g-g_{NA})^2\right)\right.\nonumber\\
&&+\left.\fr{g_1^2}{4}((N-1)\P^2+\P_N^2-N\xi)+\fr{g_2^2}{2}\fr{1}{N}(\P^2-\P_N^2)\right]\P\nonumber\\
&&\P_N''+\fr{1}{r}\P_N'=\left[\fr{1}{N^2}\left(\fr{1}{r^2}(f+(N-1)f_{NA})^2-(g+(N-1)g_{NA})^2\right)\right.\nonumber\\
&&+\left.\fr{g_1^2}{4}((N-1)\P^2+\P_N^2-N\xi)+\fr{g_2^2}{2}\fr{1-N}{N}(\P^2-\P_N^2)\right]\P_N
\label{eqs}
\eeqn

These functions obey the following
boundary conditions:
\beqn
&& \phi_{N}(0)=0,\;\;\;\phi'(0)=0\,,
\nonumber\\[2mm]
&& f_{NA}(0)=1,\;\;\;f(0)=1\,,
\nonumber\\[2mm]
&& g'(0)=0,\;\;\;g_{NA}'(0)=0\,,
\label{bc0}
\eeqn
at $r=0$, and
\beqn
&& \phi_{N}(\infty)=\sqrt{\xi_{eff}},\;\;\;\phi(\infty)=\sqrt{\xi_{eff}}\,,
\nonumber\\[2mm]
&& f_{NA}(\infty)=g_{NA}(\infty)=0,\;\;\;\; \; f(\infty) = 0,\;\;\;\;
g(\infty)=\mu N\,,
\label{bcinfty}
\eeqn
at $r=\infty$.
The boundary conditions for the derivatives were added in order to exclude
solutions that behave as logarithms at $r=0$.
The tension of this  string can be calculated by substituting the expressions
for the fields into the
energy functional, and at large $\xi_{eff}$ it behaves
as $\sqrt{\xi_{eff}}$.

To obtain the non-Abelian string solution from the $Z_N$ string
(\ref{znstr}) we apply the diagonal color-flavor rotation  preserving
the vacuum (\ref{diagphi}). To this end
it is convenient to pass to the singular gauge where the scalar fields have
no winding at infinity, while the string flux comes from the vicinity of
the origin. In this gauge we have
\beqn
\Phi &=&
U\left(
\begin{array}{cccc}
\phi(r) & 0& ... & 0\\[2mm]
.....&...&...&...\\[2mm]
0& ... & \phi(r)&  0\\[2mm]
0 & 0& ... & \phi_{N}(r)
\end{array}
\right)U^{-1}\, ,
\nonumber\\[5mm]
A^{{\rm SU}(N)}_i &=&
\frac{1}{N} \,U\left(
\begin{array}{cccc}
1 & ... & 0 & 0\\[2mm]
.....&...&...&...\\[2mm]
0&  ... & 1 & 0\\[2mm]
0 & 0& ... & -(N-1)
\end{array}
\right)U^{-1}\, \left( \pt_i \alpha\right)  f_{NA}(r)\, ,
\nonumber\\[5mm]
A^{{\rm SU}(N)}_0 &=&
-\frac1N g_{NA}\left(
\begin{array}{cccc}
1 & ... & 0 & 0\\[2mm]
.....&...&...&...\\[2mm]
0&  ... & 1 & 0\\[2mm]
0 & 0& ... & -(N-1)
\end{array}
\right)\, ,
\nonumber\\[5mm]
A^{{\rm U}(1)}_i &=& -\sqrt{\frac{2}{N}}\,
\left( \pt_i \alpha\right)   f(r)\, , \nonumber\\[2mm]
A^{{\rm U}(1)}_0 &=& \sqrt\frac{2}{N}g\,,
\label{nastr}
\eeqn
where $U$ is a matrix $\in {\rm SU}(N)$. This matrix parameterizes
orientational zero modes of the string associated with flux rotation
in  SU($N$).  The orientational
zero modes of a non-Abelian string were first
observed in \cite{auzzi,tong1}. Note that the nonabelian
strings are not BPS objects and their stability is ensured
by topological arguments. Moreover it is clear from
the solutions to the equations of motion that they
are charged with respect to  U(1) field.

\section{SUSY model}
In this Section we shall consider the vacuum structure of the SUSY theory
with nontrivial chemical potential. In what follows we shall focus on the
softly broken $\ntwo$ SUSY QCD with $N_f=N_c$ flavors.
The gauge group is SU(2)$\times$U(1) with different couplings
for abelian and nonabelian parts and we
add FI term for the $U(1)$ factor.

\beqn
{\cal L}&=&
-\frac1{4g^2_2}\left(F^{a}_{\mu\nu}\right)^2 -
\frac1{4g^2_1}\left(F_{\mu\nu}\right)^2
+\frac1{g^2_2}\left|D_{\mu}a^a\right|^2
+\frac1{g^2_1}\left|\partial_{\mu}a\right|^2
\nonumber\\[3mm]
&+& \left|\nabla_{\mu}q^{A}\right|^2 + \left|\nabla_{\mu}
\bar{\tilde{q}}^{A}\right|^2
\nonumber\\[3mm]
&+&J_0A_0
\nonumber\\[3mm]
&-& \frac{g^2_2}{2}\left( \frac{1}{g^2_2}\,  \varepsilon^{abc} \bar a^b a^c +
\bar{q}_A\,\frac{\tau^a}{2} q^A
- \tilde{q}_A \frac{\tau^a}{2}\,\bar{\tilde{q}}^A\right)^2
\nonumber\\[3mm]
&-& \frac{g^2_1}{8}\left(\bar{q}_A q^A - \tilde{q}_A\bar{\tilde{q}}^A\right)^2
\nonumber\\[3mm]
&-& \frac{g^2_2}{2}\left| \tilde{q}_A\tau^a q^A\right|^2-\frac{g^2_1}{2}\left|
\tilde{q}_A q^A - \xi_F \right|^2
\nonumber\\[3mm]
&-&\frac12\sum_{A=1}^2 \left\{ \left|(a+\sqrt{2}m_A +\tau^a a^a)q^A\right|^2
+ \left|(a+\sqrt{2}m_A +\tau^a a^a)\bar{\tilde{q}}_A\right|^2 \right\}\, + {\rm
fermions}
\label{mamodel}
\eeqn

The fields $A_\mu,\,a$ and $A^a_\mu,\,a^a$ belong to abelian and non-abelian
${\cal N}=2$ gauge supermultiplets respectively,
the fields $q^{kA}$ and $\tilde{q}_{Ak}$ represent the matter hypermultiplets.
Here $k=1,\,2$ is a color index
and $A=1,\,2$ is a flavor index. The induced chemical potential for U(1)
charge is introduced by adding the source term $A_0J_0$.

To determine the vacuum of the model consider the  bosonic potential
which in the absence of gauge fields looks as follows

\beqn
V&=&\fr{g^2_2}{2}\left(\fr{1}{g_2^2}\epsilon^{abc}\bar a^b
a^c+\bq\frac{\tau^a}{2}q^A-
\tq\frac{\tau^a}{2}\btq\right)^2+\fr{g_2^2}{2}\left|\tq\tau^a
q^A\right|^2\nonumber\\
&+&\left|(a+\sqrt{2}m_A+\tau^a a^a)q^A\right|^2+\left|(a+\sqrt{2}m_A+\tau^a
a^a)\btq\right|^2\nonumber\\
&+&\fr{g_1^2}{8}(\bq q^A-\tq\btq)^2+\fr{g_1^2}{2}\left|\tq
q^A-\xi\right|^2-\frac{\mu^2}{4}(\bq q^A+\tq\btq).\label{V}
\eeqn
Here we stated explicitly the term with vev of $A_0$ that comes from the
kinetic terms for squarks.
Let us minimize the expression in the third line $V_1$:
\beq
V_1=\fr{g_1^2}{8}(\bq q^A-\tq\btq)^2+\fr{g_1^2}{2}\left|\tq
q^A-\xi\right|^2-\frac{\mu^2}{4}(\bq q^A+\tq\btq)~.
\eeq
Upon differentiation over $q^{kB}$ and $\bar{\tilde q}^{kB}$,
we get
\beqn
\left[\fr{g_1^2}{4}(\bq q^A-\tq\btq)-\frac{\mu^2}{4}\right]\bar
q_{Bk}=-\fr{g_1^2}{2}(\bq\btq-\xi)\tilde q_{Bk}\nonumber\\
\fr{g_1^2}{2}(\tq q^A-\xi)\bar q_{Bk}=\left[\fr{g_1^2}{4}(\bq
q^A-\tq\btq)+\frac{\mu^2}{4}\right]\tilde q_{Bk}~.\label{ur1}
\eeqn
It is clear that $\bar q_{Ak}$ and $\tilde q_{Ak}$
are proportional at minimum.
Let
$q^{kA}=\fr{1}{\sqrt{2}}q\cdot A^{kA},~\bar{\tilde
q}^{kA}=\fr{1}{\sqrt{2}}\bar{\tilde q}\cdot A^{kA}$, where $q$ and $\bar{\tilde
q}$~---~complex variables and
\beq
{\rm Tr~}A^\dagger A=2.
\eeq

\beqn
\left[\fr{g_1^2}{4}(\bar q q-\tilde q\bar{\tilde
q})-\frac{\mu^2}{4}\right]\bar
q=-\fr{g_1^2}{2}(\bar q\bar{\tilde q}-\xi)\tilde q\nonumber\\
\fr{g_1^2}{2}(\tilde q q-\xi)\bar q=\left[\fr{g_1^2}{4}(\bar q q-\tilde
q\bar{\tilde q})+\frac{\mu^2}{4}\right]\tilde q~.\label{ur2}
\eeqn

\beq
(q-\tilde q)\cdot(q^2+\tilde q^2+2\xi-\fr{\mu^2}{g_1^2})=0.
\eeq
At small $\mu$ there is unique real solution $q=\tilde q$ and  we derive
\beq
q=\tilde q=\sqrt{\xi+\fr{\mu^2}{2g_1^2}}~.
\eeq
Another solution to (\ref{ur1})  $q=\tilde q=0$,
does not provide minimum.
If $m_1=m_2$, with unitary $A$ , and $a^a=0,~a=-\sqrt{2}m$,
we get absolute minimum of $V$ while
in the case $m_1\ne m_2$,
\beq
A=1,~a^3=\fr{m_2-m_1}{\sqrt 2},~ a=-\fr{m_1+m_2}{\sqrt 2}.
\eeq
When $\mu$ is large enough the second real solution to the extremum
equation emerges. Namely $q=-\tilde q$ is possible however
a quick inspection shows that it corresponds to the metastable state, provided
that $\xi\ne 0$.
The diagonalization
of the matrix of the quadratic fluctuations yields
all positive eigenvalues that is we consider the
minimum of the potential indeed.

The case of $\xi=0$ needs for a special care. The point is that
for $\xi=0$ the two branches of the solution join to a vacuum valley
\beq
q=e^{i\B}\bar{\tilde q}=\sqrt{\fr{\mu^2}{2g_1^2}}\,,
\eeq
where $\B$ is an arbitrary constant phase.
If the $\xi\neq 0$ there is nontrivial sin-Gordon potential  for this moduli
field however at large $\xi$ it is frozen at the minimum of the potential and
becomes
non-dynamical. At small non-vanishing $\xi$ this field should be taken into
account
in the low-energy approximation.

Substituting the equation $q^{kA}=\bar{\tilde q^{kA}}$ into the Lagrangian one
can see that its bosonic part
coincides with the Lagrangian of our non-SUSY model for $N=2$ and
$\Phi=\sqrt{2}q$ provided that the quark masses are equal.
So the solution for string can be expressed in terms of the profile functions

\beqn
q^{kA} &=&
\fr{1}{\sqrt 2}
U\left(
\begin{array}{cc}
\phi(r) & 0\\[2mm]
0 & \phi_{N}(r)
\end{array}
\right)U^{-1}\, ,
\nonumber\\[5mm]
A^{{\rm SU}(2)}_i &=&
\frac{1}{2} \,U\left(
\begin{array}{cc}
1 & 0\\[2mm]
0 & -1
\end{array}
\right)U^{-1}\, \left( \pt_i \alpha\right)  f_{NA}(r)\, ,
\nonumber\\[5mm]
A^{{\rm SU}(2)}_0 &=&
-\frac12 g_{NA}(r)\left(
\begin{array}{cc}
1 & 0\\[2mm]
0 & -1
\end{array}
\right)\, ,
\nonumber\\[5mm]
A^{{\rm U}(1)}_i &=&
-\left( \pt_i \alpha\right)   f(r)\, , \nonumber\\[2mm]
A^{{\rm U}(1)}_0 &=& g(r)\,,
\label{nastr}
\eeqn
which satisfy the equations (\ref{eqs}) for
$N=2$\,,

\beqn
&&\left(d_r^2+\fr{1}{r}d_r-\fr{1}{4r^2}(f-f_{NA})^2+\fr{1}{4}(g-g_{NA})^2
-\fr{g_1^2}{4}(\phi^2+\phi_N^2-2\xi)-\fr{g_2^2}{4}(\phi^2-\phi_N^2)\right)\phi=0\,,\nonumber\\
&&\left(d_r^2+\fr{1}{r}d_r-\fr{1}{4r^2}(f+f_{NA})^2+\fr{1}{4}(g+g_{NA})^2
-\fr{g_1^2}{4}(\phi^2+\phi_N^2-2\xi)+\fr{g_2^2}{4}(\phi^2-\phi_N^2)\right)\phi_N=0\,,\nonumber\\
&&d_r\left(\fr{f'}{r}\right)=\fr{g_1^2}{2r}\left(\P^2(f-f_{NA})+\P_N^2(f+f_{NA})\right)\,,\nonumber\\
&&d_r\left(\fr{f_{NA}'}{r}\right)=\fr{g_2^2}{2r}\left(-\P^2(f-f_{NA})+\P_N^2(f+f_{NA})\right)\,,\nonumber\\
&&\left(d_r^2+\fr{1}{r}d_r\right)g_{NA}=\fr{g_2^2}{2}\left(-\P^2(g-g_{NA})+\P_N^2(g+g_{NA})\right)\,,\nonumber\\
&&\left(d_r^2+\fr{1}{r}d_r\right)g=\fr{g_1^2}{2}\left(\P^2(g-g_{NA})+\P_N^2(g+g_{NA})\right)+g_1^2J_0\,.
\eeqn

Instead of matrix U it will be more convenient to use moduli $n^a$ defined as
follows,
\beq
U\tau^3U^{-1}=\vec{n}\vec{\tau}\,,\;\;\vec{n}\vec{\tau}\equiv \sum_{a}
n^a\tau^a\,,
\eeq
where $\tau^a$ are Pauli matrices.
Then the string solution can be written as follows

\beqn
&&q=\bar{\tilde
q}=\fr{1}{2\sqrt{2}}(\phi(r)+\phi_N(r)+(\phi(r)-\phi_N(r))(\vec{n}\vec{\tau}))\,,\nonumber\\
&&A_i=\Ep_{ij}\fr{x_j}{r^2}f(r)\,,\nonumber\\
&&A_i^a=-n^a\Ep_{ij}\fr{x_j}{r^2}f_{NA}(r)\,,\nonumber\\
&&A_0=g(r)\,,\nonumber\\
&&A_0^a=-n^ag_{NA}(r)\,.
\eeqn

The equation for the profile functions can be solved numerically
while the topological argument providing the stability is
the same both for SUSY and non-SUSY cases. Let us comment
on the central charges in the theory with the chemical potential.
The BPS nonabelian string saturates the stringy central
charge \cite{gs} in $N=1$ theory with matter. If chemical
potential is added the canonical momentum gets modified
and the anticommutators of supercharges are modified
as well. It can be shown that the nonabelian
string in this theory does not saturate the modified
central charge hence the equations of motion can
not be reduced to the first order ones.

\section{Worldsheet theory}

In this Section  we shall comment on the worldsheet theory of the nonabelian
string in a dense matter. To get the worldsheet Lagrangian we consider the nonabelian moduli
$n^a$ as functions of $x_k=(t, z)$ and substitute nonabelian string solution
into the initial microscopic Lagrangian , following the standard procedure
\cite{sy2004}.
The substitution for fields looks like
\beqn
&&q=\fr{1}{2\sqrt{2}}(\phi_1(r)+\phi_2(r)+(\phi_1(r)-\phi_2(r))(\vec{n}\vec{\tau}))\,,\nonumber\\
&&\bar{\tilde q}=e^{i\B}q\,,\nonumber\\
&&A_i=\Ep_{ij}\fr{x_j}{r^2}f(r)\,,\nonumber\\
&&A_i^a=-n^a\Ep_{ij}\fr{x_j}{r^2}f_{NA}(r)\,,\nonumber\\
&&A_0=g(r)\,,\nonumber\\
&&A_0^a=-n^ag_{NA}(r)-\rho\Ep^{abc}n^b\pt_0n^c\,,
A_3^a=-\rho\Ep^{abc}n^b\pt_3n^c\,.
\eeqn
Here we also want to take in account the quasimodulus $\B(x_{\mu})$ which
becomes a {\it bona fide}
vacuum modulus at $\xi=0$. The function $\rho(r)$ is an auxiliary field that
should be later eliminated from
the Lagrangian by its equation of motion.

After some calculations, we arrive at the $CP^1$ sigma model for $n^a$ fields
perturbed by a term that is actually
proportional to the chemical potential and a sine-Gordon theory for $\B$

\beqn
{\cal L}_2&=&\fr12((\pt_0n)^2-(\pt_3n)^2)\cdot\left[\fr{2\pi}{g_2^2}
\int r{\rm d}r
\left(\fr{f_{NA}^2}{r^2}(1-\rho)^2+(\rho')^2+\right.\right.\nonumber\\
&&\left.\left.g_2^2((\P^2+\P_N^2)\fr{\rho^2}{2}+(\P-\P_N)^2(1-\rho))\right)\right]\nonumber\\
&&+\fr12(\pt_3n)^2\cdot \fr{2\pi}{g_2^2}\left[\int r{\rm d}r
g_{NA}^2(1-\rho)^2\right]\nonumber\\
&&+\fr{1}{2}\int2\pi r{\rm
d}r\left[(\pt_{\mu}\B)^2-g_1^2\xi\cos\B\right]\cdot\left[(\phi_1^2+\phi_2^2)\right]\nonumber\\
&&+\fr{i}{2}\int2\pi r{\rm
d}r~\pt_0\B\cdot\left[g(\phi_1^2+\phi_2^2)-g_{NA}(\phi_1^2-\phi_2^2)\right]\nonumber\\
&&-\fr{i}{2}\int{\rm
d}^2x~\pt_i\B\cdot\Ep_{ij}\fr{x_j}{r^2}\left[f(\phi_1^2+\phi_2^2)-f_{NA}(\phi_1^2-\phi_2^2)\right]\,,
\eeqn
where $\rho (r)$ should be substituted from its equation of motion.

Let us emphasize the difference between the orientational moduli in the
worldsheet
action and $\beta$ dependent terms. The orientational moduli are purely
two-dimensional
fields while $\beta$-field is essentially four-dimensional. We have written
this
field in the worldsheet action implying that it is projected on the worldsheet but in general situation  2d integral should be substituted
by 4d space-time integral. In other words the quanta of $\beta$ field which
have mass of order $g_1\sqrt{\xi}$ can escape string worldsheet and propagate in
the bulk. They have the natural interpretation as the superpartners
of the Higgsed photon.

The two-dimensional $CP(N-1)$ model is
an effective low-energy theory relevant for the description of
internal string dynamics  at low energies,  much lower than the
inverse thickness of the string which, in turn, is given by
$g_2\sqrt{\xi_{eff}}$.
Thus,
$g_2\sqrt{\xi_{eff}}$ plays the role of a physical ultraviolet  cutoff.
and
\beq
\Lambda^N_{CP(N-1)} = g_2^N\, \xi_{eff}^{N/2} \,\, e^{-\frac{8\pi^2}{g^2_2}} .
\label{lambdasig}
\eeq
Note that in the bulk theory, due to the VEV's of
the squark fields, the coupling constant is frozen at
$g_2\sqrt{\xi_{eff}}$.

The worldsheet theory is non-supersymmetric $\sigma$-model which has
single vacuum state and which spectrum consists of kink-antikink
bound states \cite{witten}. This claim is certainly true
at small chemical potential when we obtain the perturbed
$\sigma$ -model. In this limit we can consider the
chemical potential as a small perturbation yielding
the corrections of the type $\frac{\mu}{\Lambda}$ which
are assumed to be small and can not strongly modify
the vacuum structure.
The bound states in this limit can be identified
with the monopole-antimonopole bound states from the four-dimensional
viewpoint. At larger values of $\mu$ such naive arguments
can not be justified and the quantum behavior of the worldsheet
theory of the nonabelian string deserves for separate investigation.

If we start with softly broken supersymmetric theory we should discuss
the impact of fermions. The fermionic degrees of freedom in the worldsheet
theory follow from the normalized fermion modes on the nonabelian string
solution. Generically there could be translational zero modes which
are superpartners of the bosonic coordinates of the string as well
of orientational modes. From the general index arguments we can
expect that the number of fermionic modes does not change when we
switch on the chemical potential. At small chemical potential the explicit expressions for these modes can be found in perturbation theory.

\section{Conclusion}

In this letter we have discussed the nonabelian
strings in the gauge theories with scalars supplemented
by the chemical potential. It is argued that
chemical potential works as FI term and  at large
$\mu$ the theory supports the semiclassical nonabelian strings
at weak coupling regime. The worldsheet theory at small
chemical potential
corresponds to the perturbed nonsupersymmetric $CP(N-1)$
model and the bound states of kink and antikinks
are the excitations in this theory. From the
four-dimensional viewpoint these are monopole-antimonopole
pairs. At large chemical potential theory supports
the nonabelian strings however  the worldsheet
theory is essentially different. In particular it is
unclear if monopole-antimonopole bound states
survive in this regime.

It would be interesting to extend the analysis
to the vacua with broken orientational symmetry and
with non-vanishing gluon condensate found recently in
\cite{miransky}. Another interesting problem concerns the
behavior of the nonabelian strings at nonzero temperature.
One could ask if monopoles in the Higgs phase
could escape string worldsheet  above some critical temperature.
In particular one could consider the case when $N_F>N_C$
and semilocal string solutions are present. The chemical
potential could be introduced for some flavour global
charges and no complications related with the Gauss law
shall emerge.

The work of A.G. was partly supported by grants RFBR-06-02-17382
and INTAS-05-1000008-7865 and V.M. by grants RFBR-06-02-17382,
INTAS-05-1000008-7865 and NSch-8065.2006.2.

\end{document}